\documentclass[a4paper]{jpconf}

\usepackage{graphicx}

\begin{document}
\title{Quantum objects in a sheaf framework}

\author{Antonina N. Fedorova and Michael G. Zeitlin}

\address{IPME RAS, St.~Petersburg,                                                        
V.O. Bolshoj pr., 61, 199178, Russia\\                                                     
http://www.ipme.ru/zeitlin.html, http://mp.ipme.ru/zeitlin.html}

\ead{zeitlin@math.ipme.ru, anton@math.ipme.ru}


\begin{abstract}
We consider some generalization of the theory of quantum states
and demonstrate that the consideration of quantum states as sheaves can provide, in principle, more deep
understanding of some well-known phenomena.
The key ingredients of the proposed construction are the families of sections of sheaves with values in
the proper category of the functional realizations of infinite-dimensional Hilbert spaces
with special (multiscale) filtrations decomposed into the (entangled) orbits generated by
actions/representations of internal hidden symmetries.
In such a way, we open a possibility
for the exact description and reinterpretation of a lot of quantum phenomena. 
\end{abstract}

\section{Introduction. Quantum states: functions vs. sheaves}

During a relative long period, it is well-known 
that there is a great difference between (at least) the mathematical 
levels of the investigation of quantum phenomena in different regions.  
At the same time, even advanced modern Mathematics cannot help us in the
final (at least practically accepted) analysis of the long standing quantum phenomena and
the final classification of a zoo of interpretations [1].
The well-known incomplete list is as follows:

{\bf (L)\ }  entanglement, measurement, wave function collapse, 
decoherence, Copenhagen interpretation, consistent histories, 
many-worlds interpretation/multiverse (MWI), 
Bohm interpretation, ensemble interpretation, (Dirac) self-interference, ``instantaneous'' quantum interaction,
hidden variables, etc. 

As a result, beyond  a lot of fundamental advanced problems at planckian scales 
we are still even unready to create 
the proper theoretical background for the reliable 
modeling and constructing of quantum devices far away from planckian scales.
It is very hard to believe that trivial simple solutions, like gaussians, can 
exhaust all variety of possible quantum states needed for the resolution of all contradictions, 
hidden inside the list {\bf (L)} mentioned above.
So, let us propose the following (physical) hypothesis:

{\bf (H1)\ } {\bf the physically reasonable really existing Quantum States cannot be described by means of functions.
Quantum state is a complex pattern which demands a set/class of functions/patches instead of one function 
for proper description and understanding}.
There is nothing unusual in {\bf (H1)\ } for physicists since Dirac's description of monopole.
All the more, there is nothing unusual for mathematicians who successfully used sheaves, germs, etc
in different areas.  
Definitely, the introducing of {\bf (H1)} causes a number of standard topics, the most important
of them  are motivations, formal (exact) definition and (at least) particular realizations.
Really, why need we to change our ideology after a century (since Planck) of success?
The answer is trivial and related to the list {\bf (L)} which 
is overcompleted with contradictions and misunderstanding after many decades of discussions.

\section{On the Route to Right Description: (Quantum) Patterns as Sheaves}

{\bf 2.1 Motivations}

{\bf 1).\ Arena for Quantum Evolution}
 
First of all, we need to divide the kinematical and dynamical features of a set of Quantum States ($QS$).  
From the formal point of view it means that one needs to consider some bundle $(X, H, H_x)$
whose sections are the so-called $|\psi>$ functions or $QS$. Here $X$ is (kinematical) space-time 
base space with the proper 
kinematical symmetry group (like Galilei or Poincare ones), $H$ is a total formal Hilbert space 
and $H(x)=H_x$ are fibers with their own internal structures and hidden symmetries. 
In addition, such a bundle has the corresponding structure
group which connects different fibers. Of course, in a very particular case 
we have the constant bundle with the trivial structure group but non-trivial fiber symmetry. 
Anyway, as we shall demonstrate later, 
it is very reasonable to provide the one-to-one correspondence between Quantum States and 
the proper sections
$$
 |QS> \ : \quad X \longrightarrow H,
 \qquad QS:\quad x \longmapsto H_x=H(x).
$$
As a result, we have, at least, three different symmetries inside this construction:
kinematical one on space-time, hidden one inside each fiber and the gauge-like 
structure group of the bundle as a whole.
It is obvious that the kinematical laws (like relativity principles) depend on the 
proper type of symmetry and are absolutely different in the base space and in 
the fibers. 
It should be noted that the functional realizations of fibers and the total space 
are very important for our aims. Roughly speaking, it can be supposed that physical effects 
depend on the type of the particular functional realization of formal (infinite dimensional) 
Hilbert space. E.g., it is impossible to use infinite smooth approximations, like gaussians, 
for the reliable modeling of chaotic/fractal phenomena. So, the part of Physics at quantum scales 
is encoded in the details of the proper functional realization.

{\bf 2).\  Localization and a Tower of Scales} 

 It is well-known that nobody can prove that gaussians (or even standard coherent states, etc) are an adequate 
and proper image for Quantum States really existing in the Nature.
We can suggest that at quantum scales other classes of functions or, 
more generally, other functional spaces (not $C^\infty$, e.g.) with the proper bases describe
the underlying physical processes.
There are two key features we are interested in. First of all, we need the best possible localization
properties 
for our trial base functions. Second, we need to take into account, in appropriate form, all contributions
from all internal hidden scales, from coarse-grained to finest ones.  
Of course, it is a hypothesis but it looks very reasonable:

{\bf (H2)\ }{\bf there is a (infinite) tower of internal scales in quantum region that 
             contributes to the really existing Quantum States and their evolution}.

So, we may suppose that the fundamental generating physical ``eigen-modes'' correspond 
to a selected functional realization 
and are localized in the best way. 
Let us note the role of the proper hidden symmetries which are responsible for
the quantum self-organization and resulting complexity.

{\bf 3).\  An Ensemble of Scales: Self-interaction}

As a result of the description above, we may have non-trivial ``interaction'' inside an infinite hierarchy 
of modes or scales. 
It resembles, in some sense, 
a sort of turbulence or intermittency. Of course, here the generating avatar is a representation theory
of hidden symmetries which create the non-trivial dynamics of this ensemble of hierarchies.

{\bf 4).\  Hidden Parameters and Hidden Symmetry}

It is well-known that symmetries generate all things (at least) in fundamental physics.
Here, we have a particular case where the generic symmetry corresponds to the internal hidden symmetry of 
the underlying
functional realization. Moreover, as it is proposed above, we have even the more complicated structure 
because we believe
that $QS$ is not a function but a sheaf. As a result, we have interaction between two different 
symmetries, namely hidden symmetry in the fiber, that corresponds 
to the internal symmetry of the functional realization, 
and the structure
``gauge'' group of a sheaf, which provides multifibers transition/dynamics.
Both these algebraic structures can be parametrized by the proper group parameters which can play the role of 
famous ``hidden variables'' introduced many decades ago. 

{\bf 5).\   MWI}

Of course, MWI or Multiverse interpretation can be covered by the structure sketched above.
Quantum States are the sections of our fundamental sheaf, so we can consider them as a collection of maps
between the patches of base space and fibers. All such maps simultaneously exist 
and, as an equivalence class, represent
the same Quantum State. We postpone the detailed description to the next Section but here let us mention that
each member of the full family can be considered as an object belonged to some fixed World.
Obviously, before
 measurement we cannot distinguish samples but after measurement we shall 
have the only copy in our hands.   

\noindent{\bf 2.2 On the Way to Definition}

The main reason to introduce sheaves as a useful instrument for the analysis 
of Quantum States is related to
their main property which allows to assign to every region $U$ in space-time  
$X$ some family $F(U)$ of algebraic or geometric objects such as functions or differential operators. 
The family can be restricted to smaller regions, and the compatible collections 
of families can be glued to give a family over larger regions, 
so it provides  connection between small and large scales, local and global data.
Informal construction is as follows. Let $X$ be the space-time base space (some topological space)
with a system of open subsets $U \subset X$, then for every $U$ and map $F$ the image $F(U)$ is some 
object with internal structure (more generally, 
$F(U)$ takes values in some category ${\bf H}$) such that
for every two open subsets, $U$ and $V$, $V \subset U$ there is the so-called 
restriction map (more generally, morphism in the category ${\bf H}$), 
$r_V,_U: F(U) \to F(V)$ (restriction morphism).
A map $F$ will be a {\bf presheaf} if restriction morphism satisfies the following properties:
{\bf (a)} for every open subset $U \subset X$, the restriction morphism $r_{U,U} : F(U) \to F(U)$ 
is the identity morphism,
{\bf (b)} if there are three open subsets $W\subset V\subset U$, then $r_{W,V} r_{V,U} = r_{W,U}$.
This property 
provides the connection or ordering of the underlying scales.
In other words, 
let $O(X)$ be the category of open sets on $X$, 
whose objects are the open sets of $X$ and whose morphisms are inclusions. 
Then a presheaf ${\bf F}$ on $X$ with values in category ${\bf H}$ is the 
contravariant functor from $O(X)$ to ${\bf H}$.
$F(U)$ is called the section of ${\bf F}$ over $U$ and we consider it as  some pre-image for
adequate Quantum State $|QS>$.
But our goal, in this direction, is a {\bf sheaf}, so we need to add two additional properties.
Let $\{U_i\}_{i\in I}$ be some family of open subsets of $X$, $U = \cup_{i \in I} U_i$.
{\bf (c)} If $\Psi_1$ and $\Psi_2$ are two elements of $F(U)$ and $r_{U_i,U}(\Psi_1)=r_{U_i,U}(\Psi_2)$
for every $U_i$, then $\Psi_1=\Psi_2$.
{\bf (d)} for every $i$ let a section $\Psi_i \in F(U_i)$.  $\{\Psi_i\}_{i \in I}$ are compatible if, 
for all $i$ and $j$, 
$r_{U_i \cap U_j, U_i}(\Psi_i) = r_{U_i \cap U_j, U_j}(\Psi_j)$. 
For every set $\{\Psi_i\}_{i \in I}$ of compatible sections on $\{U_i\}_{i \in I}$, 
there exists the unique section $\Psi \in F(U)$ such that $r_{U_i,U}(\Psi) = \Psi_i$ 
for every ${i \in I}$.
The section ${\bf \Psi}$ is called the gluing of the sections ${\Psi_i}$.
Definitely, we can consider this property as allusion to the hypothesis 
of wave function collapse. 
Really, ${\bf \Psi}$ looks as Multiverse Quantum State Ensemble $\{\Psi_i\}$ while $\Psi_i$ 
is the result of
measurement in the patch $U_i$. And it is unique!
The next step is to specify the Quantum Category ${\bf H}$. According to our Hypothesis ${\bf H2}$,
we consider the category of the functional realization of (infinite-dimensional) Hilbert spaces provided with
proper filtration, which allows to take into account multiscale decomposition for all dynamical quantities
needed for the description of Quantum Evolution. The well-known type of such filtration is the 
so-called multiresolution decomposition. 
It should be noted that the whole description is much more complicated because it demands the
consideration of both structures together, namely, the fiber structure generated by internal hidden symmetry 
of the chosen functional realization 
and the family of gluing sections ${\bf \Psi}$ in the unified framework.

\noindent{\bf 2.3 Realization via Multiresolution: Dynamics, Measurement, Decoherence, etc.}

In the companion paper, we shall consider in details 
one important realization of this construction based on the 
local nonlinear harmonic analysis which has, as the key ingredient, the so-called 
Multiresolution Analysis (MRA). 
It allows us to describe internal hidden dynamics on a tower of scales.
Introducing the Fock-like space structure on the whole space of internal hidden scales,
we have the following MRA decomposition:
$$
H=\bigoplus_i\bigotimes_n H^n_i
$$
for the set of n-partial Wigner functions (states):
$$
W^i=\{W^i_0,W^i_1(x_1;t),\dots,\qquad
W^i_N(x_1,\dots,x_N;t),\dots\}.
$$

So, qualitatively, {\bf Quantum Objects} can be represented by an infinite or sufficiently large set of coexisting and
interacting subsets 
while {\bf (Quasi)Classical Objects} can be described by one or a few only levels of resolution
with (almost) suppressed interscale self-interaction.
It is possible to consider Wigner functions as some measure of the quantum character of the system:
as soon as it becomes positive, we arrive to classical regime and so there is 
no need to consider the full hierarchy decomposition in the MRA representation. 
So, Dirac's self-interference is nothing else than the multiscale mixture/intermittency.
Certainly, the degree of this self-interaction leads to different qualitative types 
of behaviour, such as localized quasiclassical states, separable, entangled, chaotic etc.
At the same time, the instantaneous quantum interaction or transmission of 
(quantum) information from Alice 
to Bob takes place not in the physical kinematical space-time but in Hilbert spaces of 
Quantum States 
in their proper functional realization where there is a different kinematic life.
As a result, on the proper orbits, we have nontrivial entangled dynamics, 
especially in contrast with its classical counterpart.

\section{Conclusions}

It seems very reasonable that there are no chances for the solution of long standing problems
and novel ones if we constraint ourselves by old routines and the old zoo of simple solutions like
gaussians, coherent states and all that.
Evidently, that even the mathematical background of regular Quantum Physics demands  
new interpretations and approaches. Let us mention only the procedures of quantization 
as a generic example. In this respect, we can hope that our sheaf extension of representation for $QS$, 
which is natural from the formal point of view, may be very productive 
for the more deep understanding of the underlying (Quantum) Physics, especially, if we consider it together
with the category of multiscale filtered functional realizations decomposed into the entangled orbits 
generated by actions
of internal hidden symmetries. In such a way, we open a possibility 
for the exact description of a lot of phenomena like entanglement and measurement, wave function collapse, 
self-interference, instantaneous quantum interaction, Multiverse, 
hidden variables, etc. [2]. In the companion paper we consider the machinery needed for 
the generation of a zoo of the complex quantum patterns during  Wigner-Weyl evolution.

\section{Perspectives: On the Route to Categorification}

Sheafification together with micrlolocalization [3] and subsequent analysis of quantum dynamics on the orbits 
in the sections with special, the so-called MRA-filtrations [4], considered in this paper and in the companion one, 
are the starting points of our attempt of Categorification Program for
Quantum Mechanics and/or General Local Quantum Field Theory [5].
In some sense, we hope on the same  breakthrough as in the golden era of Algebraic Topology and Algebraic Geometry
in the 50s and 60s of the 20th Century, which was concluded by Grothendieck approach [6] and provided the universal description
for a variety of long standing problems.
Roughly speaking, such an approach provides useful, constructive and universal methods to glue the complex local data
into the general picture by power machinery taking into account the topology 
and geometry of the underlying hidden internal structures.
Definitely, the simple linear algebra of structureless Hilbert spaces cannot describe 
the whole rich world of quantum phenomena. Our approach introduces Grothendieck schemes [7] instead 
of varieties/manifolds as generic quantum objects, naturally encoded 
the full zoo of phenomenological things discussed in Quantum Mechanics.
The key ingredient of such an approach is the bridge between the von Neumann description of measurement
together with the Gelfand ideal of the state and GNS (Gelfand-Naimark-Sigal)-construction [5], [8] on one side of the river  
and locally ringed space, structure scheaf and (affine) scheme on the opposite (categorificated) side.
We will consider all technical details in the separate paper.

\end{document}